\documentclass{elsarticle}
\usepackage[utf8]{inputenc}
\usepackage{color}
\usepackage[usenames,dvipsnames]{xcolor}

\usepackage{centernot}
\usepackage{hhline}
\usepackage{mdframed}

\usepackage{amsthm}
\usepackage{bm} % for bold symbols

\usepackage[scr=boondoxo]{mathalfa} % mathscr usa il font boondoxo (calligrafico)

\usepackage{siunitx}
\usepackage{textcomp}
\usepackage{caption}
\usepackage{pst-func}
\usepackage{amssymb}
\usepackage{graphicx}
\usepackage{paralist}
\usepackage{environ}
\usepackage{pgfplots}
\usepackage{tikz}
\usetikzlibrary{shadows}
\usepackage{mathtools}
\usepackage{array,multirow}
\usepackage{algorithm}
\usepackage{pgfplotstable}
\usepackage{threeparttable}
\usepackage{extarrows}

\usepackage{etoolbox}
\usepackage{arydshln}

\usepackage{xspace} % Smart spaces after commands
 
\usepackage{dirtytalk} % for \say command
\usepackage{stmaryrd} % for llbracket and rrbracket\textbf{\textbf{}}

\usepackage[inline,shortlabels]{enumitem} 
\newlist{inlinelist}{enumerate*}{1}
\setlist*[inlinelist,1]{%
  label=(\roman*),
}

\usepackage{xifthen}  % Extended conditional commands
\usepackage{ifthen}

\usepackage{nicefrac}

\usepackage[hidelinks]{hyperref}
\usepackage{cleveref}
\usepackage{bigints}

\usetikzlibrary{pgfplots.dateplot}
\usetikzlibrary{matrix,chains,positioning,decorations.pathreplacing}
\usetikzlibrary{patterns,calc,fit,arrows}
\usetikzlibrary{shapes.geometric}

\usepackage[draft,nomargin,inline,index]{fixme} % Simplified management of FIXME's
\fxusetheme{color}

\FXRegisterAuthor{bart}{anbart}{\color{magenta} {\underline{bart}}}
\FXRegisterAuthor{ric}{anric}{\color{red} {\underline{ric}}}
\FXRegisterAuthor{zun}{anzun}{\color{OliveGreen} {\underline{zun}}}

\usepackage{listings}
\definecolor{listingBG}{HTML}{FFFFCB}%
\definecolor{listingFrame}{HTML}{BBBB98}%
\definecolor{listingLineno}{rgb}{0.5,0.5,1.0}%
\definecolor{LightGrey}{rgb}{0.975,0.975,0.975}

\lstdefinelanguage{balzac}{
	commentstyle=\color{Gray},
	morecomment=[l]{//},
	morecomment=[s]{/*}{*/},
	classoffset=0,
        escapechar=\$,
	morekeywords={const,transaction,input,output,absLock,relLock,key,network,package},
	keywordstyle=\color{TealBlue}\bfseries,
	classoffset=1,
	morekeywords={sig,versig,ctxo,rtxo,ptxo,stxo,verscr,verctx,txid,fun,unit,int,string,bool,address,uint,date,checkDate,sha256},
	keywordstyle=\color{Blue}\bfseries,
	classoffset=2,
	morekeywords={BTC,true,false,and,or,if,then,else},
	keywordstyle=\color{Plum}\bfseries,
	classoffset=3,
	morekeywords={arg,val,wit,out,in,tkid,op,owner,tkval},
	keywordstyle=\color{MidnightBlue}\bfseries,
        frame=lines,
}

\lstdefinelanguage{utxolang}{
	commentstyle=\color{Gray},
	morecomment=[l]{//},
	morecomment=[s]{/*}{*/},
	classoffset=0,
        escapechar=\$,
	morekeywords={if,then,else,@after,@pre,@receive,@next,@def,@auth,@afterRel,contract,pay},
	keywordstyle=\color{Plum}\bfseries,
	classoffset=1,
	morekeywords={balance,and,or,not,old_bal},
	keywordstyle=\color{NavyBlue},
	classoffset=2,
	morekeywords={int,string,bool,address,uint},
	keywordstyle=\color{MidnightBlue}\bfseries,
	basicstyle=\fontseries{m}\normalsize\ttfamily
	\lst@ifdisplaystyle\scriptsize\fi,
}

\newcommand{\ifempty}[3]{%
  \ifthenelse{\isempty{#1}}{#2}{#3}%
}

\newcommand{\ifdots}[3]{%
  \ifthenelse{\equal{#1}{...}}{#2}{#3}%
}

\newcommand{\hidden}[1]{}

\newcommand{\Real}[1]{\mathrm{Real}}

\newcommand{\eg}{e.g.\@\xspace}
\newcommand{\ie}{i.e.\@\xspace}
\newcommand{\wrt}{w.r.t.\@\xspace}

\theoremstyle{plain}% default

\theoremstyle{definition}
\newtheorem{defn}{Definition}

  {}

\newcommand{\sig}[3][]{\mathit{sig}^{#1}_{#2}\ifempty{#3}{}{({#3})}}

\newcommand{\BTC}{\textup{%
  \leavevmode
  \vtop{\offinterlineskip %\bfseries
    \setbox0=\hbox{B}%
    \setbox2=\hbox to\wd0{\hfil\hskip-.03em
    \vrule height .3ex width .15ex\hskip .08em
    \vrule height .3ex width .15ex\hfil}
    \vbox{\copy2\box0}\box2}}\xspace}

\def\pmvColor{\color{ForestGreen}}
\newcommand{\pmvFmt}[1]{{\pmvColor{\sf #1}}}

\newcommand{\pmv}[2][]{\pmvFmt{#2}_{\pmvColor{#1}}\xspace}
\newcommand{\pmvA}[1][]{\pmv[{#1}]{A}}
\newcommand{\pmvB}[1][]{\pmv[{#1}]{B}}
\newcommand{\pmvC}[1][]{\pmv[{#1}]{C}}
\newcommand{\pmvP}[1][]{\pmv[{#1}]{P}}
\def\txColor{\color{MidnightBlue}}
\def\fieldColor{\color{Plum}}
\newcommand{\txFmt}[1]{{\txColor{\sf #1}}}

\newcommand{\tx}[2][]{\txFmt{#2}_{\txColor{#1}}}
\newcommand{\txT}[1][]{\tx[#1]{T}}

\newcommand{\txTi}[1][]{\txFmt{T'_{\txColor{{\it #1}}}}}

\newcommand{\txTag}[3][]{{\fieldColor\sf #3}\ifempty{#1}{\ifempty{#2}{}{: {#2}}}{({#1})\ifempty{#2}{}{: {#2}}}}

\newcommand{\txIn}[2][]{\txTag[{#1}]{#2}{in}}
\newcommand{\txWit}[2][]{\txTag[{#1}]{#2}{wit}}
\newcommand{\txOut}[2][]{\txTag[{#1}]{#2}{out}}

\newcommand{\txAfterRel}[2][]{\txTag[{#1}]{#2}{relLock}}
\DeclareMathAlphabet{\mathbfsf}{\encodingdefault}{\sfdefault}{bx}{n}

\definecolor{LightGrey}{rgb}{0.95,0.95,0.95}
\definecolor{keyword}{HTML}{7F0055}

\newlength\replength
\newcommand\repfrac{.1}

\newcommand\rulewidth{.6pt}
\newcommand\tdashfill[1][\repfrac]{\cleaders\hbox to \replength{%
  \smash{\rule[\arraystretch\ht\strutbox]{\repfrac\replength}{\rulewidth}}}\hfill}

\newcommand\tdotfill[1][\repfrac]{\cleaders\hbox to \replength{%
  \smash{\raisebox{\arraystretch\dimexpr\ht\strutbox-.1ex\relax}{.}}}\hfill}

\newcommand{\var}[2][]{#2_{#1}} % variables
\newcommand{\varX}[1][]{\var[#1]{x}}

\newcommand{\varSig}[1][]{\var[#1]{\varsigma}} %change me

\newcommand{\val}[2][]{#2_{#1}} % values
\newcommand{\valV}[1][]{\val[#1]{v}}
\newcommand{\valVi}[1][]{\val[#1]{v'}}
\newcommand{\valVii}[1][]{\val[#1]{v''}}

\newcommand{\versigName}{{\sf versig}}
\newcommand{\versig}[2]{\versigName({#1},{#2})}

\newcommand{\hashSem}[1]{\ifempty{#1}{H}{H(#1)}}
\def\contrColor{\color{RubineRed}}
\newcommand{\contrFmt}[1]{{\contrColor{\it #1}}}
\newcommand{\contrC}[1][]{\mathord{\contrFmt{C}_{\contrColor{#1}}}}

\newcommand{\contrAdvC}[2]{\mathcal{C}} % computational contract advertisement

\usepackage{tikz-cd}

\newcommand{\rev}[1]{ {\it rev{:} \ {#1}} }
\newcommand{\wait}[1]{ {\it wait{:}\ {#1}} }
\tikzcdset{every label/.append style = {font = \tiny, sloped}}
\tikzcdset{every node/.append style = {font = \scriptsize}}
\tikzcdset{every arrow/.append style = {thick, swap , -Straight Barb}}
\usepackage{subcaption}
\begin{document}
	
	\title{How To Save Fees in Bitcoin Smart Contracts:\\ a Simple Optimistic Off-chain Protocol}
	
	\author{Dario Maddaloni}
	\author{Riccardo Marchesin}
	\author{Roberto Zunino} 
	\address{Università degli Studi di Trento, Italy}
	\date{21 Oct 2024}

	\begin{abstract}
		We consider the execution of smart contracts on Bitcoin.
		There, every contract step corresponds to appending to the blockchain a new transaction that spends the output representing the old contract state, creating a new one for the updated state.
		This standard procedure requires the contract participants to pay transaction fees for every execution step.
		
		In this paper, we introduce a protocol that moves most of the execution of a Bitcoin contract off-chain.
		When all participants follow this protocol, they are able to save on transaction fees, drastically reducing them.
		By contrast, whenever adversaries try to disrupt the off-chain execution, any honest participant is still able to enforce the correct contract behaviour, by continuing its execution on-chain.
	\end{abstract}
	
	\maketitle
	
	\section{Introduction} 
	
	Bitcoin was the first blockchain network, and since its creation it has been mainly used for transferring cryptocurrency in a decentralized, autonomous way.
	Bitcoin transactions however can do more than basic currency transfers by leveraging the bitcoin scripting language.
	This language is not Turing-complete, it is loop-free, it only features a few cryptographic primitives, and is in general quite limited in what it can do.
	In spite of these limitations, Bitcoin still makes it possible to implement many smart contracts. 
	While the expressiveness of Bitcoin as a smart contract platform does not reach the one of blockchains specifically designed for smart contracts (such as Ethereum),
	the class of contracts that can be implemented in Bitcoin is 
	relatively broad~\cite{BCZ18isola, BZ18bitml}.
	Some simple contracts can be expressed using a \emph{single} Pay-to-Script-Hash transaction that encodes a suitable spending condition.
	However, it is also possible to exploit \emph{multiple} transactions to implement more complex Bitcoin contracts, allowing multiple rounds of interaction among participants \cite{Andrychowicz14sp,bitcoinsok, FGCS25ComparativeAnalysis}. 
	In this work we refer to these multi-transaction protocols as ``Bitcoin smart contracts''.
	
	Users who want to deploy such a smart contract on Bitcoin can do so with the following standard technique (which we will later detail in~\Cref{subsec:on-chain-contracts}): first, they generate a tree of transactions that describes all the possible runs of the contract, then they sign all the transactions in the tree, and finally they append to the blockchain only the transactions in a single tree path, according to the participants' choices.
	While this approach is feasible, it has a main downside: every contract execution step corresponds to appending a transaction from the tree to the blockchain, which requires paying transaction fees.
	This can be expensive, especially when the number of contract steps is large.
	
	In order to improve fee efficiency, one can attempt to devise methods to perform the execution \emph{off-chain}.
	Indeed, moving the computation off-chain to save on fees has become a popular means to scale the blockchain smart contracts processing on several platforms~\cite{TSA22rollupsurvey}. 
	To this aim, a variety of methods has been considered, in particular over account-based blockchain such as Ethereum~\cite{Kalodner2018ArbitrumSP, Li19OnOffChain, DeSalve23L2DART}.

	In this work, we instead focus on saving fees in Bitcoin smart contracts.
	We propose a protocol to move most of the execution off-chain, while still guaranteeing the same contract behaviour (\Cref{sec:off-chain-contracts}).
	In this protocol participants simulate the contract by exchanging off-chain signatures. In the best case, in which all participants are honest and follow the protocol correctly, they only need to append three transactions to the blockchain. This does not depend on the number of transactions in the original contract. 
	
	Our off-chain protocol has a failsafe mechanism that can be triggered whenever someone detects malicious behaviour. 
	This mechanism moves the contract execution back on-chain, safeguarding the contract from malicious actors. Even in this negative scenario, the steps that were completed off-chain so far are preserved: the on-chain execution starts from the last off-chain state.
	This reduces the amount of on-chain steps needed to get the contract to completion.
	When the failsafe is triggered, participants need to pay the fees associated with the remaining contract steps and wait for a few additional time delays. The fees that are saved by the off-chain execution (whether full or partial) can be distributed to the participants when the contract terminates: this serves as an incentive for participants to correctly execute the off-chain protocol.
	
	While our technique is designed to be executed on Bitcoin as-is, it only relies on the fundamentals of the UTxO model, as well as timelocks, a widely adopted primitive in UTxO blockchains.
	This makes our approach easily adaptable to other UTxO platforms.
	
	\paragraph{Contributions}
	We summarize our contributions as follows:
	\begin{itemize}
		\item We consider on-chain Bitcoin smart contracts, whose execution involves appending a transaction at each step, and 
		we design a optimistic \emph{protocol} to execute such contracts off-chain.
		%
		% Our protocol is optimistic: it is able to save fees when participants are honest, but ensures contract security even in the presence of attacks.
		
		\item Our protocol can be run on \emph{stock} Bitcoin --- we do not rely on any extensions to the blockchain (\eg, new opcodes, new signature types).
		
		\item We study the \emph{security} of our protocol.
		The off-chain execution of a contract is always faithful to its on-chain one, even in the presence of adversaries.
		Further, off-chain execution steps are \emph{final}: after a honest participant completes a step, no adversary can cause the contract state to be rolled back to an old one.
		
		\item We assess the \emph{efficiency} of our protocol.
		In the best case scenario, where all participants are honest, only three transactions are needed to execute the whole contract, even when it requires a large number of steps.
		This drastically reduces the amount of fees that have to be paid to run a contract.
		In the worst case scenario, where participants misbehave right after the contract stipulation, our protocol requires to put on the blockchain two additional transactions \wrt the on-chain execution.
		Consequently, fees can only increase by a small amount.
		
		\item We showcase our technique through a simple example.
		
	\end{itemize}

    \subsection{Overview}
    
    We consider contract trees $\contrC$, whose nodes are transactions $\txT[0], \txT[1], \ldots$. Each transaction redeems its parent in the tree, except for the root $\txT[0]$ which redeems those outputs used to provide the initial contract funding. Transactions in the leaves terminate the contract, transferring its balance to participants. Tree paths from the root to leaves, represent all the possible contract execution traces, and the branches in internal nodes represent choices that participants take when the contract is executed.
    Contract trees can be executed on-chain by appending to the blockchain the transactions in a tree path, one by one, paying fees at each step.

    We then propose a protocol to move the computation off-chain, saving fees. The key idea is to put an initial transaction $\tx{Head}$ on-chain, and give all the participants an off-chain copy $\contrC[0]$ of tree $\contrC$, called a {\em graft}, that can redeem $\tx{Head}$.
    In principle, any participant can now force the execution to be moved back on-chain by starting to append transactions from $\contrC[0]$. However, participants can also decide which is the next contract step to make $\txT[0] \to \txT[i]$, so identifying the subtree $\contrC[i]$ of $\contrC[0]$ which is rooted at $\txT[i]$. Participants are then given a new graft, a copy of $\contrC[i]$ which redeems $\tx{Head}$, and they could now move the execution on-chain by appending the transactions from $\contrC[i]$, if they so wish.
    Participants can instead continue the off-chain execution by agreeing on the next step $\txT[i] \to \txT[j]$, and sharing a graft for $\contrC[j]$ rooted at $\txT[j]$.
    When a leaf is reached, appending its related graft on-chain causes the contract to directly move into its final state without having to pay the fees for each intermediate step.

    Our protocol features a few additional mechanisms to protect the off-chain execution from misbehaving participants. In particular, Bitcoin timelocks and an intermediate transaction $\tx{Init}$ are exploited so to give newer grafts higher priority over older ones and avoid state rollbacks.

	\section{Preliminaries}
    
    \subsection{On-chain contracts}\label{subsec:on-chain-contracts}
	
	As mentioned in the introduction, we want to consider Bitcoin contract that execute across multiple transactions. We formalize this concept below as a \emph{contract tree}.
	Intuitively, the nodes of the tree will represent the possible states of the contract, while the edges give a set of conditions that must be satisfied in order to pass from a state to another.
	
	To run Bitcoin contracts on-chain, we follow a technique inspired from~\cite{BZ18bitml,BCZ18isola}.
	Participants must first follow the \emph{stipulation} protocol (\Cref{def:on-chain-stipulation-protocol}), which commits them to actually execute the contract later on.
	After that, participants must follow the actual \emph{execution} protocol (\Cref{def:on-chain-execution-protocol}).

	\paragraph{Handling fees in Bitcoin contracts}
	%	By choosing an off-chain execution technique for contracts we want to be able to reduce the amount of fees payed by the participants.
	Bitcoin requires participants to pay a fee each time a transaction is appended to the blockchain.
	The price of fees varies according to different factors, such as the congestion of the Bitcoin network and the size of the transaction.
	The (on chain) execution of Bitcoin smart contracts requires appending multiple transactions to the blockchain, and that needs fees to be paid.
	To ensure that the contract execution can always progress, we must guarantee that the contract always holds enough funds to pay the required fees.
	For that reason, we require contract participants to provide enough currency in advance, during the contract \emph{stipulation phase}, so that the contract always can pay fees without needing more funds during the \emph{execution phase}.
	
	During the stipulation phase, we assume participants can reasonably predict the needed fees.
	For the sake of keeping our presentation simple, in our model we denote with $\sf fee$ a fixed Bitcoin value which is adequate for paying the fees of any contract transaction.
	
	We can now formalize on-chain Bitcoin contract trees as follows.
	
	\begin{defn}[Contract tree]\label{def:contract-tree}
		A \emph{contract tree} $\contrC$ is a rooted tree of Bitcoin transactions satisfying the requirements below.
		We distinguish among three types of contract nodes:
		\begin{itemize}
			\item The \emph{root} $\txT[0]$, that may have any number of inputs, but only has a single output.
			\item The \emph{inner nodes}, that have a single input and a single output.
			\item The \emph{leaves}, that have a single input and may have multiple outputs.
		\end{itemize}
		
		Contract transactions must preserve currency: the cumulative value of transaction outputs must be equal to the cumulative value of its inputs, minus the value of a $\sf fee$.
		
		Except for $\txT[0]$, the inputs of contract transactions refer to outputs of other contract transactions.
		Instead, the inputs of $\txT[0]$ refer to outputs of transactions \emph{outside} $\contrC$.
		We require that any such output can only be redeemed with the signature of a participant. We will refer to such participants as the \emph{contract participants}.
		
		The edges of $\contrC$ are determined by the input-output dependencies of the transactions, \ie there is an edge from $\txT[i]$ to $\txT[j]$ if and only if the input of $\txT[j]$ refers to the output of $\txT[i]$. Edges are labelled by zero or more \emph{unlock conditions}, which are enforced by transaction scripts (to be explained below).
		\qed
	\end{defn}
	
	In our contract, we allow the following unlocking conditions:
	
	\begin{defn}[Unlock conditions]
		The conditions that can label a contract edge from $\txT[i]$ to $\txT[j]$ are as follows:
		\begin{itemize}
			\item Authorization: the condition $sig_{\pmvA}$ requires contract participant $\pmvA$ to sign $\txT[j]$, authorizing the operation.
			\item Hash decommitting: the condition $\rev{S}$ requires to reveal the secret $S$ whose hash $H(S)$ was previously committed by a contract participant.
			\item Waiting: the condition $\wait{t}$ requires waiting for at least $t \cdot \Delta_{w}$ seconds after $\txT[i]$ before appending $\txT[j]$. We fix the time granularity $\Delta_{w}$ to be one hour. We require $t>0$.
			\qed
		\end{itemize}
	\end{defn}
	
	Our choice of $\Delta_{w}$ is one hour because that is the time needed for a transaction to be confirmed in the Bitcoin blockchain. Conventionally, this happens after six blocks are appended on the blockchain.
	While different values of the granularity $\Delta_{w}$ could be used, we will assume that $\Delta_{w}$ is large enough to allow contract participants to run off-chain protocols in a timely fashion, with no risk of making $\wait{t}$ expire.
	More precisely, exchanging protocol messages among participants must take very little time compared to $\Delta_{w}$.
	Further, adversaries that aim to delay other participants by being unresponsive when communicating with them must be detected before a $\wait{t}$ expires.
	For that, we fix a time $\Delta_{r} \ll \Delta_{w}$, and we will require that honest protocol participants detect when another participant whose message is expected takes more than $\Delta_{r}$ to reply.
	$\Delta_{r}$ must still be large enough to allow honest participants to compute and send their messages.
	
	We now discuss the transaction scripts that must occur in contract trees. These enforce the unlocking conditions on the tree edges.
	
	\begin{defn}[Contract transaction scripts]
		\label{def:contract-scripts}
		The script of a non-leaf transaction $\txT[i]$ has the form
		$scr_{j_1} \lor \cdots \lor scr_{j_n}$ where $\txT[j_1], \cdots \txT[j_n]$ are the children of $\txT[i]$ in the contract tree.
		Further $scr_{j}$ performs the following checks:
		\begin{itemize}
			\item All participants must have signed the redeeming transaction, certifying that it is indeed $\txT[j]$.
			For this task each participant uses a distinct key for each $j$.
			We will refer to these signatures as the \emph{implicit signatures}, since they are required no matter what the unlocking conditions are.
			\item For each $sig_{\pmvA}$ label on the $(i,j)$ edge, the participant $\pmvA$ must have provided an additional signature with a different key from the above ones.
			\item For each $\rev{S}$ label, a preimage of $H(S)$ must be included among the witnesses of $\txT[j]$, effectively making $S$ public.
			\item For each $\wait{t}$ label, $\txT[j]$ must include a relative timelock that prevents it to be appended to the blockchain before $t$ units of time since $\txT[i]$.
			\qed
		\end{itemize}
		
	\end{defn}

	\begin{figure}
		\begin{center}
			\centering
			\begin{tikzcd}
				\colorbox{yellow}{$\tx[\pmvA]{Dep}$} \arrow[rrd, dotted, "sig_{\pmvA}"] &  &                                                                                                            & {\tx[1]{T}} &                                                                        & {\tx[4]{T}} \\
				\colorbox{yellow}{$\tx[\pmvB]{Dep}$} \arrow[rr, dotted,"sig_{\pmvB}"]  &  & {\tx[0]{T}} \arrow[rr, "{sig_{\pmvB}, \rev{S_{\pmvA}}}"] \arrow[rd, "{sig_{\pmvC}}"] \arrow[ru, "\wait{5}"] &             & {\tx[2]{T}} \arrow[ru, "{sig_{\pmvA,\pmvB}}"] \arrow[rd, "\wait{10}"] &             \\
				\colorbox{yellow}{$\tx[\pmvC]{Dep}$} \arrow[rru, dotted, "sig_{\pmvC}"] &  &                                                                                                            & {\tx[3]{T}} &                                                                        & {\tx[5]{T}}
			\end{tikzcd}
			\caption{Example of a contract tree. 
				$sig_{\pmvP}$ labels require an authorization by $\pmvP$ (signature on the redeeming transaction).
				$\rev{S_{\pmvA}}$ labels require revealing the secret $S_{\pmvA}$ whose hash was previously committed.
				$\wait{t}$ labels require waiting for at least $t \cdot \Delta_w$ time.
			}
			\label{fig:contract:example}
		\end{center}
	\end{figure}
	
	An example of a contract tree $\contrC$ is depicted in \Cref{fig:contract:example}.
	There, we show a few ``deposit'' transactions which are already on the blockchain, highlighted in yellow, which are not part of $\contrC$.
	The root $\txT[0]$ of $\contrC$ refers to such transactions as its inputs, as denoted with the dotted lines, which we label with the signatures which are required to spend them.
	$\pmvA, \pmvB,\cdots$ denote the contract participants.
	
	% Edges in the tree represent requirements that need to be satisfied in order for a transaction to be redeemed. Such restrictions are either set by the script of the \say{parent} transaction (for instance asking to sign the redeeming transaction with a specific key, or to reveal the preimage of a given hash), or they can be set within the redeeming transaction itself, like in the case of timelocks, that force the transaction to wait for a certain amount of blocks before it can be appended. 
	
	The rest of the transactions form the actual contract tree $\contrC$.
	We represent the edges of $\contrC$ with solid arrows, corresponding to the scripts as defined in~\Cref{def:contract-scripts}.
	We label the solid lines with their unlocking conditions.
	We remark that transaction scripts enable such edges only when the condition is satisfied, but they also require an \emph{implicit} signature by each contract participant.
	In all our figures, we do not show such implicit signatures on solid lines so to avoid clutter.
	
	\paragraph{On contract leaves}
	We remark that or definition of contract tree essentially poses no requirement on the outputs of the contract leaves.
	Typically, leaves will distribute the contract balance in some way among participants, but they can also transfer Bitcoins to other parties, or even require complex redeeming conditions.
	For instance, in the example above, the contract is designed so that its leaves transfer the bets to the winning participant. 
	More in general, the actual script used in contract leaves is mostly immaterial to our results.
	However, as we will see in~\Cref{subsec:more-on-fees}, it is possible to design leaves so to redistribute the unspent transaction fees which arise from executing a short contract path.
	This will be taken into account when assessing the cost of executing a contract (both in the on-chain and off-chain cases).
	
	\subsection{Stipulation and Execution of On-chain Contracts}
	
	\paragraph{Stipulation}
	Contract stipulation involves the following steps:
	
	\begin{defn}[On-chain stipulation protocol]\label{def:on-chain-stipulation-protocol}
		\mbox{}
		\begin{enumerate}
			\item Participants must first communicate their interest in stipulating $\contrC$, making it known to everyone.
			
			\item Each participant then generates all the transactions in the contract tree, except for their witnesses.
			
			\item Then, participants exchange all the \emph{implicit} signatures (as per~\Cref{def:contract-scripts}) on each transaction. If an invalid implicit signature is received, the stipulation is aborted.%
			\footnote{
				We stress that these implicit signatures do not include those needed to satisfy the authorizations in the unlock conditions. Those will be only be made during the contract execution.}
			
			\item Participants exchange the signatures for the root transaction $\txT[0]$, so that it can redeem its input.
			
			\item $\txT[0]$ is put on the blockchain. The contract is now \emph{stipulated}.
			\qed
		\end{enumerate}
	\end{defn}
	
	The last item, appending $\txT[0]$, is the ``point of no return''.
	Stipulation could be aborted at any time before that, but once $\txT[0]$ is on the blockchain the contract must now be executed.
	
	An important aspect of the stipulation protocol is that all the implicit signatures in the contract tree are exchanged and verified before $\txT[0]$ is signed.
	If we allowed $\txT[0]$ to be signed earlier, it would be possible to put it on the blockchain and start the execution of the contract before all the implicit signatures are made, allowing an adversary to deny some of their implicit signature and effectively stall the execution of the contract, freezing its funds forever.
	
	\paragraph{Execution}
	Once the stipulation protocol is concluded and $\txT[0]$ is on the blockchain, participants can execute the contract by suitably appending the contract transactions along any tree path, as long as the unlocking conditions in that path are satisfied.
	This is formalized by the following protocol:
	
	\begin{defn}[On-chain execution protocol]
		\label{def:on-chain-execution-protocol}
		\mbox{}
		
		\begin{enumerate}
			
			\item Let $\txT[*]$ be the single transaction of $\contrC$ which is on the blockchain with its outputs still unspent.
			
			\item If $\txT[*]$ is a leaf of $\contrC$, the execution of the contract is over.
			
			\item \label{def:on-chain-exe:3} If $\txT[*]$ is a non-leaf, we consider its children. Participants may choose to reveal their secrets, to exchange their signatures on some child transactions, or to wait, according to their wishes. In this way, they can enable the transition from $\txT[*]$ to the children they wish to put on the blockchain.
			
			\item \label{def:on-chain-exe:4} Any participant can then choose one of the enabled children $\txT[j]$ and append it to the blockchain, redeeming $\txT[*]$ in the process.
			We stress that the script of $\txT[*]$ is satisfied: all the implicit signatures have already been exchanged in the stipulation phase, while the unlocking conditions have been satisfied in the previous step (\Cref{def:on-chain-exe:3}).
			
			\item After $\txT[j]$ has been appended to the blockchain, we restart the protocol from the beginning. \qed
		\end{enumerate}    
	\end{defn}
	
	Note that in \Cref{def:on-chain-exe:3} the protocol does \emph{not} force participants to reveal their secrets or exchange signatures.
	Rather, honest participants can freely choose whether to do so.
	For instance, if a contract tree node has two children and both require an authorization form participant $\pmvA$, then the contract effectively grants $\pmvA$ to choose, at execution time, which contract subtree must be executed.
	Here, $\pmvA$ could authorize either branch, both, or none.
	
	Further, we remark that in~\Cref{def:on-chain-exe:4} there might be multiple children that can be appended on the blockchain.
	If that happens, multiple (honest) participants can attempt appending different transactions on the blockchain, all spending $\txT[*]$.
	This causes a race: the Bitcoin network will effectively choose which transaction is appended, and which is discarded.
	This is not an issue on its own. If undesired, the contract can be designed to reduce such races by carefully using the unlocking conditions, including temporal constraints, so that one of the options becomes enabled before the other, making it possible to avoid the race.
	
	We remark that the execution protocol only requires a single participant to be honest.
	Indeed, the presence of a single honest participant is enough to guarantee that the contract will be executed according to one of its paths.
	This is because each non-root node requires an implicit signature by \emph{all} participants, and a honest participant does not reuse the key they used in the stipulation protocol to sign new transactions. 
	Consequently, only the stipulated paths can be followed.
	
	Further, the execution protocol does not require participants to interact except when
	a participant wants to satisfy an unlocking condition, and has to reveal secrets or share new signatures.
	A single honest participant is still able to make the contract progress on its own, without any help from others, as long as they choose a path whose unlocking conditions can be satisfied by them.
	Indeed, as already remarked in~\Cref{def:on-chain-exe:4}, the implicit signatures are already known by the honest participant.
	The contract execution can therefore only stop when reaching a contract leaf, therefore exiting the contract itself, or when reaching an internal node whose children all require authorizations or secrets which are being refused.
	The last case can be avoided by designing the contract so that each non leaf always has a children requiring at most a $\wait{t}$ condition: in this way a leaf can always be reached.

	\subsection{Example: On-chain Best-of-3 Bet} \label{subsec:bet-example}
	We illustrate the on-chain execution of a contract through an example.
	The tree of transactions presented in \Cref{fig:Bo3Bet} implements a best-of-three bet between Alice and Bob, on some kind of recurring match between two teams. 
	Before the competition starts, an oracle commits the hashes of six secret messages: three of them denote that the first team won a match, while the other three denote that it lost%
	\footnote{The 6 messages contain the strings $W_1, W_2, W_3, L_1, L_2, L_3$ and are padded with a nonce for security. Note that the oracle is not a contract participant, and does not need to be aware of the bet between Alice and Bob. Indeed its role could be performed by three distinct oracles, one for each match.
	}.
	After each match the oracle certifies the victory or loss of the team by revealing the suitable secret message. These secrets can then be used by Alice and Bob to update the state of the contract, appending a new transaction to the blockchain. 
	Transactions in the contract are named to represent the win-loss record associated to the corresponding state, from the point of view of the first team, meaning that $\tx{W??}$ represents the state in which the first team has won once, and two matches still have to be played, while $\tx{LWL}$ represents the state in which the first team has lost, then won, then lost again.
	The contract ends when either team has won two out of three matches, or when both Alice and Bob agree to an early payout, terminating the bet before the three matches have been played. The contract has 10 possible terminations, shown in the tree as 10 leaves.
	
	The transactions $\tx{WLW}, \tx{WW}$ transfers all the balance of the contract to Alice, who bet on the first team winning. Likewise, the transactions $\tx{LL}, \tx{LWL}$ transfer the prize to Bob. The early payout option, that can be taken only if Alice and Bob both agree on it, is performed by leaves $\tx[W]{Out}$ (which gives a bigger share to Alice, since it can be taken after the first win has been revealed), $\tx[WL]{Out}$ and $\tx[LW]{Out}$ (which split the prize equally), and $\tx[L]{Out}$ (which pays a bigger share to Bob). 
	
	It is important to notice that in this contract the closer a leaf is to the root, the more total money it will have available, due to having to pay less in fees. For this reason, the transactions $\tx{WW}, \tx{LL}, \tx[W]{Out}, \tx[L]{Out}$ can pay back to Alice and Bob a part of the fees that they anticipated.
	In such a small contract, the difference is only slight, but in deeper contracts, \ie a best-of-5 bet, they start to become more impactful.
	\Cref{fig:Bo3Bet:path} shows a possible on-chain execution of the contract.
	
	\begin{figure}
		\begin{center}
			\centering
			% https://tikzcd.yichuanshen.de/#N4Igdg9gJgpgziAXAbVABwnAlgFyxMJZABgBoAmAXVJADcBDAGwFcYkQAdDnAD2S7QBbWgEFKwACIw0AXxAzS6TLnyEUZACzU6TVuy69+HIbQBC4qbPmKQGbHgJEAjKQDM2hizaJO3HsFMYHDkFJXtVInIKD11vX15gAHUAfmSQmzsVRxQorRpPPR8DfwAZVPSwrLVkV1JiGK99P2RE8QB5ZmDrSodq2qp82KaExMSK22VeolqnBsL4-0Sy8cyplFq8nUaiv2ASlJXJiPW3ObjivZLD8Oya0gBWM+G+EvbO66qiDVJZwe2FpJLD5rZDfAZbeYXJZXboTG7Vb71P6Q5pLN5dUJwz4ob7uZHnXb7Maw1bHUGkTYFAkJfYwzGk27fR7457IfbokLaGBQADm8CIoAAZgAnCCCJAuEA4CBIciYkVipBkKUyxBymwK8VqmjSpC1ECMegAIxgjAACkdsiAsGBsLAQCydsKYLQkk45DQ4AALLCCnBK+WirVRFVIb4G42mi3w9g2u1sR2+Z2ukru2GavU61X3GiGk3my1qA0wP0OiFxbA8gD6wAEwhECgABJWa3WzOMM4h9brEAA2XORgsxnzCrA8r3+xNcZNJOXpoOZ0OIADsA-z0ex1ttWHtU44M5Kc8DisQ4Z7AA411HC+xGCXJ+WmjPEnPPT7S4hiMetWfVQBOK8h03OMdwTR8nRdPYjw1BdTyzCVlTza9hy3eMyypdgW1rYx6wULC21MDtYNXJcnElJCgLWEBR3HB8MIglNXCIk8SJ7JwQwojcqJA3dwKTSDEiY+cWPgxAnH1Tibx8HiwPokB8Jw0QmwUkxCOErVL1I8NJJQmT0KGBikiE78kE0ticwjdcpOLUs9wPYyYJPMzVScftLOQ4Dt14uSVNw0hmzHVtFLUmRKBkIA
			\begin{tikzcd}
				&                                                                                &                                                                                                                       & {\tx[W]{Out}}                                                                                                           & {\tx[WL]{Out}} \\
				&                                                                                &                                                                                                                       & \tx{WL?} \arrow[r, "\rev{L3}"] \arrow[rd, "\rev{W3}"] \arrow[ru, "{sig_{\pmvA, \pmvB}}"] & \tx{WLL}       \\
				\colorbox{yellow}{$\tx[\pmvA]{Dep}$} \arrow[rd, dotted, "sig_{\pmvA}"] &                                                                                & \tx{W??} \arrow[ruu, "{sig_{\pmvA,\pmvB}}"] \arrow[r, "\rev{W2}"] \arrow[ru, "\rev{L2}"]          & \tx{WW}                                                                                                                 & \tx{WLW}       \\
				& \tx{Bet} \arrow[ru, "\rev{W1}"] \arrow[rd, "\rev{L1}" ] &                                                                                                                       & \tx{LL}                                                                                                                 & \tx{LWW}       \\
				\colorbox{yellow}{$\tx[\pmvB]{Dep}$} \arrow[ru, dotted, "sig_{\pmvB}"] &                                                                                & \tx{L??} \arrow[r, "\rev{W2}"] \arrow[ru, "\rev{L2}" ] \arrow[rd, "{sig_{\pmvA, \pmvB}}"] & \tx{LW?} \arrow[ru, "\rev{W3}"] \arrow[r, "\rev{L3}"] \arrow[rd, "{sig_{\pmvA, \pmvB}}" ]  & \tx{LWL}       \\
				&                                                                                &                                                                                                                       & {\tx[L]{Out}}                                                                                                           & {\tx[LW]{Out}}
			\end{tikzcd}
		\end{center}
		\caption {The Best-of-3 bet contract tree crafted during stipulation. Only the deposit transactions are on the blockchain at this time (highlighted in yellow).
			The stipulation phase completes by putting $\tx{Bet}$ on chain, after which execution can start.
		}
		\label{fig:Bo3Bet}
	\end{figure}
	
	\begin{figure}
		\begin{center}
			\centering
			\begin{tikzcd}
				&                                                                                &                                                                                                                       & {\tx[W]{Out}}                                                                                                           & {\tx[WL]{Out}} \\
				&                                                                                &                                                                                                                       & \tx{WL?} \arrow[r, "\rev{L3}"] \arrow[rd, "\rev{W3}"] \arrow[ru, "{sig_{\pmvA, \pmvB}}"] & \tx{WLL}       \\
				\colorbox{yellow}{$\tx[\pmvA]{Dep}$} \arrow[rd, dotted, "sig_{\pmvA}"] &                                                                                & \tx{W??} \arrow[ruu, "{sig_{\pmvA,\pmvB}}"] \arrow[r, "\rev{W2}"] \arrow[ru, "\rev{L2}"]          & \tx{WW}                                                                                                                 & \tx{WLW}       \\
				& \colorbox{yellow}{$\tx{Bet}$} \arrow[ru, "\rev{W1}"] \arrow[rd, "\textcolor{black}{\rev{L1}}", yellow, very thick] &                                                                                                                       & \tx{LL}                                                                                                                 & \tx{LWW}     \\
				\colorbox{yellow}{$\tx[\pmvB]{Dep}$} \arrow[ru, dotted, "sig_{\pmvB}"] &                                                                                & \colorbox{yellow}{$\tx{L??}$} \arrow[ru, "\rev{L2}"] \arrow[r, "\textcolor{black}{\rev{W2}}", yellow, very thick ] \arrow[rd, "{sig_{\pmvA, \pmvB}}"] & \colorbox{yellow}{$\tx{LW?}$} \arrow[ru, "\rev{W3}"] \arrow[r, "\textcolor{black}{\rev{L3}}", yellow, very thick] \arrow[rd, "{sig_{\pmvA, \pmvB}}" ]  & \colorbox{yellow}{$\tx{LWL}$}       \\
				&                                                                                &                                                                                                                       & {\tx[L]{Out}}                                                                                                           & {\tx[LW]{Out}}
			\end{tikzcd}
		\end{center}
		\caption {A possible contract execution path. After the contract is fully executed, all the transactions highlighted in yellow are on the blockchain, while the others became invalid and must be kept off-chain (or deleted).
			To perform the steps in the path, participants must wait for the oracle to de-commit one of the hashes, so that they exploit the corresponding preimage to redeem the previous transaction in the path with the next one, so moving to the next contract state.}
		\label{fig:Bo3Bet:path}
	\end{figure}
	
	\subsection{More on fees}
	\label{subsec:more-on-fees}
	
	For the sake of simplicity, in~\Cref{def:contract-tree} we used a fixed amount of currency ($\sf fee$) for the fees of any transaction.
	We remark that this simplification is not strictly necessary, and that different fees can be used throughout the contract tree, \eg, making them to be proportional to the transaction size.
	We still require that the inputs on any tree node are sufficient to pay for the fees, hence the contract root $\txT[0]$ must obtain from its inputs enough currency to pay for all the needed fees in any of its execution paths.
	
	Note that different tree execution paths might use a different amount of fees, whether because the individual transactions fees are different, or simply because a tree path is shorter, hence it requires fewer execution steps.
	Because of this, the contract balance which becomes available to tree leaves can vary.
	While not strictly mandatory, in most common contracts the root transaction requires all the contract participants to equally share the fees cost.
	Dually, the contract leaves refund the unspent fees back to the participants.
	In the cheaper execution paths, the unspent fees can be significant, while in the most expensive  paths there are no unspent fees to refund.

	\section{Off-chain contracts}\label{sec:off-chain-contracts}
	
	In this section we present a technique to move the contract \emph{off-chain}.
	For this, we use new off-chain stipulation and execution protocols.
	Before describing them in a general way (Definitions~\ref{def:off-chain-stipulation-protocol} and \ref{def:off-chain-execution-protocol}), we show them in the specific case of the best-of-three bet example of~\Cref{subsec:bet-example}.
	
	\subsection{Example: off-chain best-of-three bet}
	We now walk through the steps that are needed to move off-chain the best-of-three bet.
	First, Alice and Bob generate a slightly modified copy of the contract (see \Cref{fig:Bo3Bet:offchain:start}), where there is a new $\tx{Head}$ transaction that has as inputs the two initial deposits, an $\tx{Init}$ transaction (whose input is $\tx{Head}$), and a modified $\tx{Bet}$ transaction, which now features a timelock of $3 \Delta_w$ relative to its input, $\tx{Init}$. From there, the modified contract follows the structure of the original one.
        % \footnote{The differences with the original contract are indeed minor ones. Transaction $\tx{Bet}$ now has $\tx{Init}$ as its only input, instead of the original deposits, and has an additional timelock.}
        
	Again, mirroring the protocol described for the on-chain stipulation of the contract, Alice and Bob send each other the implicit signatures needed for every transaction, leaving for last the ones that unlock the two deposits and enable $\tx{Head}$.
	Once they have all the needed signatures, they append $\tx{Head}$ to the blockchain, spending their deposits.
	Now the on-chain transactions are exactly those highlighted in~\Cref{fig:Bo3Bet:offchain:start}.
	Notice that, at this point, the transaction $\tx{Init}$ is enabled (since they had already exchanged the implicit signatures) but, unlike in the on-chain contract execution, Alice and Bob should refrain from appending it to the blockchain right away.
	Indeed, appending $\tx{Init}$ would interrupt the off-chain execution of the contract, forcing it to be executed on-chain, and causing Alice and Bob to pay the fees at each step. 
	The ideal scenario is the one in which $\tx{Init}$ is appended only after the off-chain execution is completed (\ie a contract leaf is reached). 
	The fact that $\tx{Init}$ is enabled from the start is an intended feature of the protocol: it is, in fact, a failsafe mechanism, that allows one participant to force the on-chain execution whenever the other is behaving maliciously and refusing to cooperate with the off-chain protocol. 
	
	From now on, our example will follow a specific execution trace of the contract, the one shown in \Cref{fig:Bo3Bet:path}, assuming that the oracle will reveal \say{L1}, \say{W2}, and \say{L3}, and that neither Alice nor Bob will agree to take an early payout.
	After the oracle reveals \say{L1}, in the original on-chain contract Bob would append $\tx{L??}$ to the blockchain moving the state forward. 
	Instead, in the off-chain protocol, the two participants create a \emph{graft}: a copy of the subtree rooted at $\tx{L??}$, modifying $\tx{L??}$ so that its input is now the $\tx{Init}$ transaction, and so that it has a relative timelock of $2 \Delta_w$. Then, they graft this subtree to the off-chain contract (see \Cref{fig:Bo3Bet:offchain:step-1}) by exchanging the implicit signatures needed by the transactions in the subtree.
	Completing this graft effectively makes the off-chain contract perform a step to the new $\tx{L??}$ state.
	
	Note that, if at this point either participant wanted to bring the contract execution on-chain, then they could do so by appending the $\tx{Init}$ transaction, waiting for $2 \Delta_w$ units of time, and then appending $\tx{L??}$, carrying on with the on-chain execution from that point onward.
	We remark that, after $\tx{Init}$ has been appended to the blockchain, an adversary could attempt to \say{roll back} the contract to a previous state, by appending $\tx{Bet}$ instead of $\tx{L??}$. However, due to the delays enforced by the timelocks, any honest participant is always able to prevent this by appending $\tx{L??}$ \emph{first}.
	Indeed, $\tx{L??}$ unlocks one $\Delta_w$ \emph{earlier} than $\tx{Bet}$.
        % ), so an honest participant can always act first.
        %: by appending $\tx{L??}$ as soon as possible they can effectively prevent the contract state to be rolled back.
	
	\begin{figure}
		\hspace{-50pt}
		\begin{minipage}{0.5\textwidth}
			\begin{tikzcd}[every label/.style = {font = \tiny}]
				{\colorbox{yellow}{$\tx[\pmvA]{Dep}$}} \arrow[rd, dotted, "sig_{\pmvA}", sloped, above] &                                          & {\colorbox{yellow}{$\tx[\pmvB]{Dep}$}} \arrow[ld, dotted, "sig_{\pmvB}", sloped, above  ]  \\
				& \colorbox{yellow}{$\tx{Head}$} \arrow[d] &    \\
				& \tx{Init} \arrow[d, "\wait{3}",right]         &    \\
				& \tx{Bet} \arrow[ld, "\rev{W1}", sloped,above] \arrow[d, "\rev{L1}", right]            &   \\
				\cdots                                            & \cdots  &  
			\end{tikzcd}
			\subcaption{Start of the off-chain tree}
			\label{fig:Bo3Bet:offchain:start}
		\end{minipage}
		\begin{minipage}{0.5\textwidth}
			% https://tikzcd.yichuanshen.de/#N4Igdg9gJgpgziAXAbVABwnAlgFyxMJZABgBpiBdUkANwEMAbAVxiRAB12BjCBiAJwBGEAB7AAnjAZ8A7gF9gAEk44RyTmgC2NAIIVgAERho5iuSDml0mXPkIoATOSq1GLNpx58hoiVNkKyuyq6uxaNABC+kYmZhZWIBjYeAREAIykaS70zKyIHNy8AsJiktIQ8koqYgASMHRQpuaW1sl26aQO2W55BarAAJJguM0JSbapKBkAzN25HsFiETA4o60T9iSkACxz7vmeUBA4CC2JNimbGbvUOfsFXEcnAATx65dE06Sztz0L-QAZAD8QLW5zak2QThurnmB24T1OYwu7RQXxhd16h2OSPeqOQ2x2eyxCJxFhcMCgAHN4ERQAAzfgQTRIMggHAQJAOM6M5lIDLszmIbkJXks4XUDlIaY8pnir6CpDbWV8xCExWIACsKvF6qliAAbDrpZKhQB2aiCGBgKCs42IC0agAc9sd+oAnK7TfziHIKHIgA
			\begin{tikzcd}[every label/.style= {font= \tiny}]
				{\colorbox{yellow}{$\tx[\pmvA]{Dep}$}} \arrow[rd, dotted,"sig_{\pmvA}", sloped, above] &                                          & {\colorbox{yellow}{$\tx[\pmvB]{Dep}$}} \arrow[ld, dotted,"sig_{\pmvB}", sloped, above] &                                          &        \\
				& \colorbox{yellow}{$\tx{Head}$} \arrow[d] &                                                   &                                          &        \\
				& \tx{Init} \arrow[d, "\wait{3}", left] \arrow[rrd, "\wait{2}", sloped, above]          &                                                   &                                          &        \\
				& \tx{Bet} \arrow[ld, "\rev{W1}", sloped, above] \arrow[d, "\rev{L1}",right]            &                                                   & \tx{L??} \arrow[ld, "sig{\pmvA, \pmvB}", sloped, above] \arrow[d, "\colorbox{white}{$\rev{W2}$}"] \arrow[rd, "\rev{L2}", sloped, above] &        \\
				\cdots                                            & \cdots                                   & \cdots                                            & \cdots                                   & \cdots
			\end{tikzcd}
			\subcaption{After the first step, a subtree is grafted}
			\label{fig:Bo3Bet:offchain:step-1}
		\end{minipage}
		\caption{}
	\end{figure}
	
	The next off-chain steps are performed in a similar fashion: Alice and Bob can graft a copy of the subtree of transactions rooted in the one that represents the next state, modifying it so that the root has as input the transaction $\tx{Init}$, and so that it now includes a timelock which is crucially smaller than the ones created in the previous steps.
	By using a smaller timelock, this graft is given a higher priority \wrt the previously created grafts.
	After the graft is created, they exchange the implicit signatures, leaving for last the ones for the subtree root. 
	
	Assuming now that Alice and Bob are continuing the off-chain execution, they will wait for the oracle to reveal the next result (in this case \say{W2}), then graft the subtree rooted at $\tx{LW?}$.
	Finally, after \say{L3} is revealed, they can graft the subtree consisting only of the transaction $\tx{LWL}$, getting to the full off-chain tree shown in \Cref{fig:Bo3Bet:offchain:full}. Now that the contract has no further possible steps, Alice and Bob complete the execution by putting both $\tx{Init}$ and $\tx{LWL}$ on-chain.
	In this way, the contract is completed with only 3 on-chain transactions ($\tx{Head}$, $\tx{Init}$, and $\tx{LWL}$), when the on-chain protocol required 4 transactions for the same execution path. Consequently, the participants have saved one fee, which can be redistributed by $\tx{LWL}$.
	
	\begin{figure}
		\hspace*{-60pt}
		% https://tikzcd.yichuanshen.de/#N4Igdg9gJgpgziAXAbVABwnAlgFyxMJZAJgBoAGAXVJADcBDAGwFcYkQAdDgYwkYgBOAIwgAPYAE8YjfgHcAvsAAkXHKORc0AW1oBBSsAAiMNPKXyQ80uky58hFABYK1Ok1bsuvfsLGTpcooqHGoaHNq0AEIGxqbmltYgGNh4BEQAzKQAjK4MLGyInDx8giLiUjIQCsqq4gASMPRQZhZWNin2GaTEue4FRWrAAJJguK2JyXZpKFmk6b35niHikTA44+1TDsizjgsehV5QEDgIbUm2qdvkpHs0eQdF3MenAAQJm1dd8-d9S4MAGQA-ECNhcOtMSLd9v0jiczhNLp0UJk7m5FoceC8EZ9kchnGiHrCsfCPuCtkQAGxzGH-cQAgDqoLJky+KGphL+mOepPOrLxAFZob8MU9sSykZCAOzC9GPOGnCUQ7YADhpIvly2AjIBrVcMCgAHN4ERQAAzAQQLRIG4gHAQJDEc4Wq1IWZ2h2IJ2JF3Wr00e1IdLOy1+zIepCOEOuxDOCOIAXRv1xwOISlJoMBz0ykBCGBgKA2jOIHOplXF0uegCcFazbvIxfDqayWWLLbriCywZ9obd7ub3vNvc7-c9WSjPZjTbHQtz+cLiHSDco8iAA
		\begin{tikzcd}[every label/.style ={font= \tiny}]
			&                               & {\colorbox{yellow}{$\tx[\pmvA]{Dep}$}} \arrow[rd, dotted, "sig_{\pmvA}", sloped, above] &                                                                        & {\colorbox{yellow}{$\tx[\pmvB]{Dep}$}} \arrow[ld, dotted, "sig_{\pmvB}", sloped, above] &        &                                          &        &          \\
			&                               &                                                   & \colorbox{yellow}{$\tx{Head}$} \arrow[d]                               &                                                   &        &                                          &        &          \\
			&                               &                                                   & \tx{Init} \arrow[lld, "\wait{3}",sloped,above] \arrow[d, "\wait{2}", right] \arrow[rrrd, "\wait{1}", sloped, above] \arrow[rrrrrd, bend left = 10] &                                                   &        &                                          &        &          \\
			& \tx{Bet} \arrow[d, "\rev{W1}", right] \arrow[ld, "\rev{L1}", sloped, above] &                                                   & \tx{L??} \arrow[ld, "sig_{\pmvA, \pmvB}", sloped, above] \arrow[d, "\colorbox{white}{$\rev{W2}$}"] \arrow[rd, "\rev{L2}", sloped, above]                               &                                                   &        & \tx{LW?} \arrow[ld, "sig_{\pmvA, \pmvB}", sloped, above] \arrow[d, "\colorbox{white}{$\rev{W3}$}"] \arrow[rd, "\rev{L3}", sloped, above] &        & \tx{LWL} \\
			\cdots  & \cdots                        & \cdots                                            & \cdots                                                                 & \cdots                                            & \cdots & \cdots                                   & \cdots &         
		\end{tikzcd}
		\caption{Complete off-chain contract}
		\label{fig:Bo3Bet:offchain:full}
	\end{figure}

	\subsection{Stipulation and Execution of Off-chain Contracts}
	
	Moving away from the example, we now present the general  off-chain stipulation and execution protocol for an arbitrary contract. 
	In the rest of this section, we denote with $\contrC$ an on-chain contract tree that we wish to execute off-chain.

    The fields of all the transactions created in this protocol are shown in~\Cref{fig:head-init-transaction,fig:graft-transactions} and discussed in~\Cref{subsec:off-chain-txs}.
    
	\paragraph{Stipulation}
	
	The off-chain stipulation protocol is similar to the on-chain one, but involves two more transactions ($\tx{Head}$ and $\tx{Init}$).
	Intuitively, the $\tx{Head}$ transaction is used to make participants commit to executing the contract.
	Instead, the $\tx{Init}$ transaction acts as a failsafe mechanism against malicious behaviour, allowing any participant to move the contract execution from off-chain to on-chain, at any time.
	
	\begin{defn}[Off-chain stipulation protocol]
		\label{def:off-chain-stipulation-protocol}
		\mbox{}    
		\begin{enumerate}
			\item Create a transaction $\tx{Head}$,
			making its inputs provide the initial contract funds, similarly to the inputs of the root of $\contrC$, but with two additional $\sf fee$s. Contract participants must provide for all these funds, as in the on-chain stipulation protocol.
            
			\item Create a transaction $\tx{Init}$ redeeming $\tx{Head}$.
			
			\item Create a copy $\contrC[0]$ of the contract tree $\contrC$, and modify it so that the root of the tree now redeems $\tx{Init}$ instead of the outputs external to the contract.
			Redeeming $\tx{Init}$ in this way must require waiting for the depth of $\contrC[0]$ times $\Delta_w$: this is implemented in Bitcoin adding a relative locktime to the root of $\contrC[0]$.
			
			\item Make the scripts of each non-leaf transaction mentioned above require a signature from every contract participant in order to be redeemed. Such signatures are made with a single-use key, much like the implicit signatures of the on-chain contract.
			We will therefore also refer to these signatures as the \emph{implicit} ones.
			The scripts in $\contrC[0]$ still check the unlocking conditions as in the original $\contrC$.
			
			\item Exchange the implicit signatures for every transaction. (Note that this does not involve signing $\tx{Head}$.)
			
			\item Sign $\tx{Head}$ and append it to the blockchain, redeeming its inputs.
			After that, the contract becomes \emph{stipulated}.
            \qed
		\end{enumerate}
	\end{defn}

	We stress that, after $\tx{Head}$ is on the blockchain, the contract could be executed on-chain by any participant.
	To do so, it suffices to 
	\begin{inlinelist}
		\item redeem $\tx{Head}$ with $\tx{Init}$, 
		\item redeem $\tx{Init}$ (after the timelock expired) with the root of $\contrC[0]$, and finally 
		\item continue from these with the on-chain contract execution.
	\end{inlinelist}
	This is possible because, after stipulation, all the implicit signatures are known by all participants.
	
	Now that participants have established this baseline security guarantee, they can attempt to execute the contract off-chain according to the following  optimistic protocol.
	
	\begin{defn}[Off-chain execution protocol]
		\label{def:off-chain-execution-protocol}
		\mbox{}    
		\begin{enumerate}
			\item Let $\contrC[*]$ be a variable referring to a copy of a subtree of $\contrC$.
			Initially, we let $\contrC[*]$ be $\contrC[0]$, the copy of $\contrC$  generated during the stipulation phase.
			
			Intuitively, $\contrC[*]$ represents the current off-chain contract state.
			
			\item\label{def:off-chain-exe:2} Run the \emph{grafting protocol} in~\Cref{def:grafting-protocol} repeatedly, aborting it immediately as soon as one of the following conditions is met:
			\begin{enumerate}
				\item The transaction $\tx{Init}$ is appended to the blockchain. (This can only happen because of a dishonest participant.)
				\item A (dishonest) participant fails to cooperate, \ie some grafting protocol message is expected from them but it is not received within time $\Delta_r$.
				\item $\contrC[*]$ becomes a tree containing only one leaf, completing the off-chain execution of the contract.
			\end{enumerate}
			
			\item\label{def:off-chain-exe:3} Append $\tx{Init}$ to the blockchain (if not already there).
			\item Append the root of the latest graft $\contrC[*]$ as soon as its timelock expires. 
			
			\item If $\contrC[*]$ comprises more nodes than just a leaf, continue the execution of the contract on-chain, appending the  transactions of $\contrC[*]$, following the on-chain execution protocol of~\Cref{def:on-chain-execution-protocol}.
            \qed
		\end{enumerate}
	\end{defn}
	
	The above protocol repeatedly creates grafts $\contrC[*]$ in~\Cref{def:off-chain-exe:2}, effectively advancing the current off-chain contract state, one step per graft.
	To do so, it uses the grafting protocol (to be defined below).
	Creating a new graft, however, requires the cooperation of all the contract participants, and as such it is subject to some threats from adversarial participants.
	We will study such threats in~\Cref{subsec:threats}.
	Here, we only remark that the off-chain execution protocol defends itself from such threats by detecting them in~\Cref{def:off-chain-exe:2}, and then by moving the execution of the contract on-chain (from \Cref{def:off-chain-exe:3}), as a failsafe mechanism.
	This is possible since, after each graft is created, participants have the option to append $\tx{Init}$ to the blockchain, wait for the root of the latest graft $\contrC[*]$ to become enabled, and append it, so that the contract can be continued from there.
	
	We now detail the protocol to create the next graft.
	
	\begin{defn}[Grafting protocol]
		\label{def:grafting-protocol}
		\mbox{}  
		In the following, we let $\txT[*]$ be the root of $\contrC[*]$, and $\txT[1],\cdots, \txT[k]$ be its children.
		\begin{enumerate}
			\item \label{def:grafting-protocol:1} Each participant share with the others their wishes about which child (or children) among $\txT[1],\cdots, \txT[k]$ should be the next contract state%
			\footnote{%
				A participant should not agree to child $\txT[j]$ when they are not intending to reveal the secrets or provide the signatures to satisfy the unlocking conditions for the step $\txT[*]\to\txT[j]$.
				Doing so will cause the grafting protocol to abort later on, eventually causing the contract execution to be moved on-chain. Even in such case, the security of the contract is not compromised.
			}.
			Participants must then reach an agreement on a single child $\txT[i]$.
			If that is not possible because of conflicting wishes, the protocol is aborted. (This makes the execution protocol continue from~\Cref{def:off-chain-exe:3}.)
			
			\item Let $\contrC[i]$ be a copy of the subtree rooted in $\txT[i]$, modified as follows:
			\begin{itemize}
				
				\item The root of $\contrC[i]$ has as its input the $\tx{Init}$ transaction (instead of $\txT[*]$).
				
				\item The root of $\contrC[i]$ uses a relative timelock that requires waiting for a time equal to the depth of $\contrC[i]$ times $\Delta_w$.
				
				\item The values of all the non-leaf outputs of $\contrC[i]$ are increased by $\sf fee$ to account for the saved fees due to the step $\txT[*] \to \txT[i]$ being now only performed off-chain.
				
				\item The values of the outputs of each leaf transaction of $\contrC[i]$ are increased, refunding the participants with the saved $\sf fee$. \footnote{We do not specify the exact redistribution method, since it is immaterial to our protocol. Participants may choose any redistribution method as long as they agree on it.}
				
				\item The keys used in the scripts in $\contrC[i]$ for verifying the implicit signatures are substituted with fresh keys, so that they keys are still single-use (they are only used to sign a single transaction), similarly to the procedure used for creating $\contrC[0]$ in the stipulation phase.
			\end{itemize}
			
			\item Participants must now satisfy the unlocking conditions of the $\txT[*]\to \txT[i]$ edge.
			This includes waiting, revealing secrets, and sending signatures. The witnesses of (the copy of) $\txT[i]$ are adapted accordingly.
			
			\item \label{protocol:sig-subtree} Participants exchange all the implicit signatures for the transactions in $\contrC[i]$. The signatures for the root are exchanged last. After the exchange is completed, we say that $\contrC[i]$ becomes \emph{grafted}.
			
			\item We set the current off-chain state $\contrC[*]$ to be $\contrC[i]$.
            \qed
		\end{enumerate}
	\end{defn}
	
	We remark that the signatures generated in \Cref{protocol:sig-subtree} are indeed needed, and the previously generated signatures cannot be reused. This is because the transactions in the grafts have different input fields from the ones in the original tree, and signatures must be calculated on the whole transaction.
	Like in the on-chain contract protocol, we emphasize that signing the whole graft before its root is essential to avoid stalling attacks. For this reason we can not sign the graft transactions step-by-step after the graft root is already on-chain: at that time all the implicit signatures must already be known by participants.

	\subsection{Detailed transaction format}
	\label{subsec:off-chain-txs}
    
    \begin{figure}
	    \centering
        \small
        \begin{tabular}{|l|}
            \hline
            \\[-9pt]
            \multicolumn{1}{|c|}{$\tx{Head}$} \\
            \hline
            \\[-9pt]
            \txIn{$\tx[\pmvA]{Dep}$.\txOut{}(0), $\tx[\pmvB]{Dep}$.\txOut{}(0) , $\cdots$ } \\ 
            \txWit{$\sig{\pmvA}{},  \sig{\pmvB}{} , \cdots$} \\
            \txOut{}(0): $ (\lambda \vec{\varSig}.
                     \versig{\pmvA{\sf Head}}{\varSig[1]}
                     \wedge \versig{\pmvB {\sf Head}}{\varSig[2]} \wedge \cdots, \valV \BTC)
                    $ \\
            \hline
        \end{tabular}

        \begin{tabular}{|l|}
            \hline
            \\[-9pt]
            \multicolumn{1}{|c|}{$\tx{Init}$} \\
            \hline
            \\[-9pt]
            \txIn{$\tx{Head}.\txOut{}(0)$} \\
            \txWit{$\sig{\pmvA {\sf Head}}{},  \sig{\pmvB {\sf Head}}{} , \cdots$} \\
              \txOut{}(0): $ ( \lambda \vec{\varSig}.
                     \versig{\pmvA{\sf Init}}{\varSig[1]}
                     \wedge \versig{\pmvB{\sf Init}}{\varSig[2]} \wedge \cdots, \valVi \BTC)  
                    $ \\
            \hline
        \end{tabular}
        
        \caption{$\tx{Head}$ and $\tx{Init}$ transactions}
	    \label{fig:head-init-transaction}
	\end{figure}

    \begin{figure}
	    \centering
        \small
        \begin{tabular}{|l|}
            \hline
            \\[-9pt]
            \multicolumn{1}{|c|}{Root $\txT[i,i]$ of graft $\contrC[i]$, copy of $\txT[i]$ with children $\txT[j_1],\txT[j_2],\ldots$} \\
            \hline
            \\[-9pt]
            \txIn{$\tx{Init}$.\txOut{}(0)} \\ 
            \txAfterRel{$ {\sf depth}(\contrC[i]) \cdot \Delta_w$}\\
            \txWit{$\sig{\pmvA {\sf Init}}{},  \sig{\pmvB {\sf Init}}{} , \cdots$} \\
            %\txOut{$ 0 \mapsto (\lambda \vec{\varSig}.
            %         \versig{\pmvA(i,i)}{\varSig[1]}
            %         \wedge \versig{\pmvB(i,i)}{\varSig[2]} \wedge \cdots, \valV \BTC)
                    % $} \\
            \txOut{}(0): $
                     (\lambda \vec{\varSig}.
                     \bigvee_k \left( \begin{array}{l}
                        \versig{\pmvA(i,j_k)}{\varSig[1]}
                        \wedge \versig{\pmvB(i,j_k)}{\varSig[2]} \wedge \cdots \wedge \\
                        \text{unlocking conditions for $\txT[i] \to \txT[j_k]$ without implicit $\sf versig$s}
                        %\text{script of $\txT[j_k].\txOut{}(0)$ without implicit $\sf versig$s}
                        \end{array}\right)
                     \, , \,
                     \valVii \BTC)
                    $ \\
            \hline
        \end{tabular}
        \begin{tabular}{|l|}
            \hline
            \\[-9pt]
            \multicolumn{1}{|c|}{Internal node $\txT[i,j]$ of graft $\contrC[i]$, copy of $\txT[j]$ child of $\txT[p]$} \\
            \hline
            \\[-9pt]
            \txIn{$\txT[i,p].\txOut{}(0)$} \\
            \txWit{$\sig{\pmvA(i,j)}{},  \sig{\pmvB(i,j)}{} , \cdots$} \\
              \txOut{}(0):  \text{analogous to the graft root output} \\
            \hline
        \end{tabular}
        \begin{tabular}{|l|}
            \hline
            \\[-9pt]
            \multicolumn{1}{|c|}{Leaf node $\txT[i,j]$ of graft $\contrC[i]$, copy of $\txT[j]$ child of $\txT[p]$} \\
            \hline
            \\[-9pt]
            \txIn{$\txT[i,p].\txOut{}(0)$} \\
            \txWit{$\sig{\pmvA(i,j)}{},  \sig{\pmvB(i,j)}{} , \cdots$} \\
            \txOut{}(0): $( \lambda \vec{\varSig}. \versig{\pmvA}{\varSig[1]} , {\valV}_{\pmvA,j} \BTC)$ \\
            \txOut{}(1):  $(\lambda \vec{\varSig}. \versig{\pmvB}{\varSig[1]} , {\valV}_{\pmvB,j} \BTC)$ \\
            $ \quad \vdots \qquad \qquad  \vdots \qquad \vdots \qquad \vdots$ \\
            \hline
        \end{tabular}

        \caption{Graft transactions}
	    \label{fig:graft-transactions}
	\end{figure}

    \begin{figure}
	    \centering
        \small
        \begin{tabular}{|l|}
            \hline
            \\[-9pt]
            \multicolumn{1}{|c|}{Graft root $\txT[Bet, Bet]$, copy of $\tx{Bet}$} \\
            \hline
            \\[-9pt]
            \txIn{$\tx{Init}.\txOut{}(0)$} \\
            \txAfterRel{$3\Delta_w$}\\
            \txWit{$\sig{\pmvA \sf Init}{},  \sig{\pmvB \sf Init}{} $} \\
            \txOut{}(0):
                $\begin{array}{l} 
                  ( \lambda \vec{\varSig} \varX.
                    (\versig{\pmvA(\tx{Bet},\tx{L??})}{\varSig[1]}
                     \wedge \versig{\pmvB(\tx{Bet},\tx{L??})}{\varSig[2]} \wedge \hashSem{\varX}= H(L1) )\vee
                  \\
                   (\versig{\pmvA(\tx{Bet},\tx{W??})}{\varSig[1]}
                     \wedge \versig{\pmvB(\tx{Bet},\tx{W??})}{\varSig[2]} \wedge \hashSem{\varX}= H(W1) ), \valV \BTC) 
                \end{array}$\!\!\!
                \\
            \hline

        \end{tabular}

        \begin{tabular}{|l|}
            \hline
            \\[-9pt]
            \multicolumn{1}{|c|}{Graft internal node $\txT[Bet,L??]$, copy of $\tx{L??}$} \\
            \hline
            \\[-9pt]
            \txIn{$\tx{Bet}.\txOut{}(0)$} \\
            \txWit{$\sig{\pmvA(\tx{Bet},\tx{L??})}{},  \sig{\pmvA(\tx{Bet},\tx{L??})}{}, L1 $} \\
            \txOut{}(0): 
                $\begin{array}{l} 
                 (
                  \lambda \vec{\varSig} \varX.
                    (\versig{\pmvA(\tx{Bet}, \tx{LL?})}{\varSig[1]}
                     \wedge \versig{\pmvB(\tx{Bet}, \tx{LL?})}{\varSig[2]} \wedge \hashSem{\varX}=H(L2) ) \vee
                  \\
                   (\versig{\pmvA(\tx{Bet},\tx{LW?})}{\varSig[1]}
                     \wedge \versig{\pmvB(\tx{Bet},\tx{LW?})}{\varSig[2]} \wedge \hashSem{\varX}=H(W2) ) \vee
                   \\
                    (\versig{\pmvA(\tx{Bet},\tx[L]{Out})}{\varSig[1]}
                     \wedge \versig{\pmvB(\tx{Bet},\tx[L]{Out})}{\varSig[2]} \wedge
                     \versig{\pmvA}{\varSig[3]}
                     \wedge \versig{\pmvB}{\varSig[4]})
                     , \valVi \BTC) 
                \end{array}$\!\!\!
                \\
            \hline
        \end{tabular}
        
        \caption{Example of graft transactions.}
	    \label{fig:graft-transaction}
	\end{figure}

    The exact format of the transactions created by our off-chain protocol are shown in~\Cref{fig:head-init-transaction,fig:graft-transactions}.
    The notation for the Bitcoin transaction in the figure is taken from~\cite{bitcointxm}.

    Our transactions involve a number of distinct cryptographic keys, which must be exchanged by the participants. More in detail, we assume that each participant $\pmvA$ has a ``master'' signing key pair whose public part is shared with other participants before contract stipulation starts.
    Each message from $\pmvA$ is then signed with that master key so to ensure its authenticity. In particular, when $\pmvA$ needs to generate new key pairs to sign certain transactions, $\pmvA$ shares the public key with the other participants, using the master key to certify that the public key is indeed $\pmvA$'s key and that its purpose is to sign a given transaction.
    As shown in the figures, each participant $\pmvA$ needs to generate key pairs $\pmvA \sf Head$ and $\pmvA \sf Init$ for the $\tx{Head}$ and $\tx{Init}$ transactions. Further, each time a graft $\contrC[i]$ is created, copying transaction $\txT[i]$ from the original contract tree and its descendants $\txT[j]$, each participant generates a key $\pmvA(i,j)$ for each transaction $\txT[i,j]$ in the graft. 
    Finally, we assume that each participant own a key to redeem basic deposits of Bitcoins. This key, which we simply denote with $\pmvA$, is only used at contract stipulation (in $\tx{Head}$) and at contract termination (in graft leaves).
    
    We now comment on the transactions and their fields.
    The $\tx{Head}$ transaction spends a few deposits owned by the participants to provide the initial contract balance $\valV\BTC$, as shown in the input field $\txIn{}$. The witness field $\txWit{}$ of the transaction carries the needed signatures. These witnesses can be segregated away according to the SegWit Bitcoin mechanism~\cite{BIP141}, if so wished. The single output $\txOut{}$ involves a script that takes a few signatures $\vec{\varsigma}$ and verifies them with the public parts of the $\pmvA {\sf Head}, \ldots$ key pairs.

    Transaction $\tx{Init}$ redeems $\tx{Head}$ in a similar way. Note that it involves its own set of keys. Its output value $\valVi \BTC$ is the value $\valV \BTC$ in $\tx{Head}$, minus a transaction fee.

    Graft transactions are more complex. 
    In the figures we show the format of the transactions corresponding to the root, internal nodes, and leaf nodes of the graft.
    The root redeems $\tx{Init}$, and uses the relative locktime field $\txAfterRel{}$ to require that $ {\sf depth}(\contrC[i]) \cdot \Delta_w$
    time has passed since $\tx{Init}$ was appended on the blockchain. The witnesses provide the required signatures.
    The script accounts for all possible children in the graft: its form is a logical disjunction of the conditions required for each move. Indeed, for each possible move $\txT[i] \to \txT[j_k]$ we require the implicit signatures made with keys $\pmvA(i,j_k), \ldots$, as well as the other specific unlocking conditions for that move (as per~\Cref{def:contract-scripts}). The script for the off-chain graft transaction $\txT[i,i]$ is therefore the same as the one used in the on-chain transaction $\txT[i]$, except that the implicit signatures are made with distinct keys each time a graft is created.
    The output value $\valVii \BTC$ is the value $\valVi \BTC$ in $\tx{Init}$, minus a transaction fee.

    The transactions for internal graft nodes are similar to the root, except for the following differences. First, they redeem their parent in the graft instead of $\tx{Init}$. Second, they are signed with different keys ($\pmvA(i,j)$ instead of $\pmvA Init$). Their output is completely analogous to the one for the graft root, and its value is once again the same as the value of the parent minus a transaction fee.
    Note that internal nodes can optionally involve a $\txAfterRel{t}$ field when there is a $\wait{t}$ unlocking condition from their parent. (By contrast, the root transition in the graft does not need to account for its own $\wait{t}$ since the waiting is to be performed off-chain. Its $\txAfterRel{}$ field is instead used to ensure it has the right priority.)

    Finally, transactions for leaf nodes simply redeem their parent and redistribute the contract balance to participants. This includes both the amount which would have been distributed by the on-chain contract \emph{and} the additional savings from the avoided fees.
    Recall from~\Cref{subsec:more-on-fees} that, even in the on-chain execution, at the end of the contract the balance can be different depending on the length of the tree path that was executed. When the contract is stipulated, the initial balance must account for all possible execution paths, their fees at each step, and their intended transfers to the participants once the contract ends. Since longer paths require more fees to be paid, shorter paths can be reached with a larger balance than desired, in which case the contract designer can choose to redistributed the unspent fees to the participants.
    
    In the off-chain contract execution, this phenomenon is also present: if a graft $\contrC[i]$ is used to move the computation on-chain, in the general case we still have paths of different lengths requires different fees.
    On top of the fees which are saved because we took a shorter path, we have the savings coming from the off-chain execution: no fees have been spent to move from the contract tree root to the graft root.
    Effectively, the number of the additional saved fees due to off-chain execution is ${\sf depth}(\contrC) - {\sf depth}(\contrC[i]) - 2$, where the number $2$ accounts for the $\tx{Head}$ and $\tx{Init}$ transactions.
    In~\Cref{fig:graft-transactions}, the values in the leaf outputs $\valV[\pmvA,j] \BTC$ sum to the contract balance in the leaf, and comprise both the intended end-of-contract transfers and the redistributed unspent fees.

    As a concrete example, we show in~\Cref{fig:graft-transaction} a few transactions from the Bet contract in~\Cref{subsec:bet-example}. These are from $\contrC[Bet]$, the first graft created by the protocol, so its root is a copy of the $\tx{Bet}$ transaction which is the root of the (on-chain) contract tree $\contrC$.
    The graft root $\txT[Bet,Bet]$ redeems $\tx{Init}$ and its script allows moves towards $\tx{L??}$ and $\tx{W??}$, depending on which hash preimage $x$ has been revealed. The relative timelock is $3\cdot \Delta_w$ since the depth of the graft is the same as the one of the whole $\contrC$ contract, i.e., we need at most three moves to reach a leaf.
    The second transaction in the figure is internal node $\txT[Bet,L??]$ which redeems $\txT[Bet,Bet]$ using the appropriate signatures and the $L1$ preimage. The script used in the output is a three-way disjunction.
    The first two cases allow moving towards $\tx{LL?}$ or $\tx{LW?}$, depending on which new preimage $x$ is revealed at this time.
    Instead, the last case allows moving towards $\tx{Out_L}$: the contract requires that this step is authorized at execution time by both participants. Therefore, as already done in the on-chain contract tree, the script requires both implicit signatures (which are exchanged at stipulation time), and additional authorization signatures (to be provided during execution)\footnote{
        This implementation would be optimized to require only the two execution-time signatures, as done in~\cite{BZ18bitml}. Here we use distinct signatures to make the purpose of each signature more evident.
    }.
    
	\subsection{Threat analysis}
	\label{subsec:threats}
	
	We now study the potential security threats that can affect the off-chain execution protocol.
	Intuitively, the safety of our off-chain protocol is predicated on three properties.
	\begin{itemize}
		\item First, we want any off-chain execution of a contract to be coherent with an on-chain execution of the same contract. 
		More in detail, each off-chain execution step must have an on-chain counterpart.
		
		\item Second, we want to ensure that any on-chain execution of a contract can be also performed off-chain.
		More precisely, whenever participants can take an on-chain step from a state, the off-chain execution protocol must allow them to take a corresponding step from the corresponding state.
		In particular, this ensures that the off-chain execution never stalls the contract except in those cases where the on-chain execution already stalls.
		
		\item Finally, we want our off-chain execution to be \emph{final}: if the off-chain protocol reaches a certain state, then the adversary must not be able to \say{roll-back} the contract to a previous state and continue the execution from there.
	\end{itemize}
	In our threat analysis we will postulate the existence of at least one honest contract participant, who is assumed to be interested in having the contract proceed exactly as stipulated.
	% As we will show, this honest participant must also be vigilant in order to thwart attacks.
	
	\paragraph{Off-chain behavior can not exceed the on-chain one}
	The off-chain execution protocol proceeds by generating and signing grafts, in this way updating the current off-chain contract state.
	By construction, honest participants sign a new graft only when the original on-chain contract allows a step between the old and new contract state and the associated unlocking conditions are met, hence no adversary can cause grafts to be produced unless following the original contract semantics.
	
	An adversary can also attempt to disrupt the contract execution by first appending $\tx{Init}$ and the root of a graft, so moving the contract execution on-chain, and then continue its attack from there.
	However, every transaction in the graft requires a signature from every participant, and one of them is honest, so the adversary can not redeem (non-leaf) graft transactions except by using other graft transactions which are linked by an graft edge.
	Since grafts are copies of subtrees of the original contract, execution can never diverge from the intended behaviour.
	
	\paragraph{Off-chain behavior includes the on-chain one}
	When all participants are honest, they are able to emulate any execution step of the on-chain contract by signing the corresponding graft.
	The required unlocking conditions are the same.
	
	In the presence of adversary, instead, it is possible that the graft can not be created even when it corresponds to an on-chain step, since the adversary can simply refuse to sign it.
	An honest participant, however, detects that the adversary is not cooperating with time $\Delta_r$, and uses the failsafe mechanism: $\tx{Init}$ is put on the blockchain, as well as the root of the most recent graft.
	Since now the execution is moved on-chain, the signatures from the adversary are no longer required.
	Indeed, all the implicit signatures have already been provided during the graft creation.
	The honest participant can therefore execute the wanted step.
	
	Note that when executing a contract on-chain it is possible that multiple (possibly honest) participants attempt to perform distinct execution steps, causing a race on the blockchain between multiple transactions that spend the same output.
	When executing the same contract off-chain, participants have several options to handle such races.
	Ideally, participants should attempt to reach an off-chain agreement on what should be the next step to take (as per~\Cref{def:grafting-protocol:1} of~\Cref{def:grafting-protocol}).
	For instance, participants could run a sub-protocol to securely generate a random number and use it to choose among the multiple options.
	If an agreement can not be reached on the next step, the execution must be moved on-chain.
	Doing so causes a race on the blockchain, exactly as in the on-chain execution.

	\paragraph{Finality of the execution}
	A more subtle adversary could attempt to disrupt the execution not by making it diverge from the contract or by stalling it, but by rolling it back to a previous state.
	
	Recall that, after off-chain stipulation, $\tx{Head}$ has been appended to the blockchain, while $\tx{Init}$ has been signed by all the participants and can be used to redeem $\tx{Init}$ at any time.
	Further, the off-chain execution protocol repeatedly sign grafts that can redeem $\tx{Init}$: one graft for each off-chain execution step.
	During this, the adversary receives a sequence of signed grafts.
	
	With this knowledge, the adversary can try to append $\tx{Init}$ and then redeem that using the root of \emph{any old graft}, effectively moving the execution of the contract on-chain while restarting it from an \emph{old} state.
	This would be problematic in several ways.
	First, the adversary could make the contract progress on-chain following contract tree path which is different from the one that has already been followed off-chain, which is undesirable.
	Second, when executing this different path, the adversary has still access to the secrets that have been revealed during the off-chain execution of the other path.
	This might be harmful to the participants which would not have wanted to reveal such secrets in this different path.
	
	To avoid such a roll-back attack, our off-chain execution protocol exploits temporal constraints on grafts.
	More precisely, grafts are created so that they can only redeem $\tx{Init}$ after a certain time has passed.
	Newer grafts are given a shorter time lock, so that they become enabled to redeem $\tx{Init}$ \emph{before} all the previously created grafts.
	Effectively, newer grafts have higher priority.
	To thwart roll-back attacks, honest participants must constantly be vigilant and detect when someone appends $\tx{Init}$ on the blockchain (\Cref{def:off-chain-exe:2} of~\Cref{def:off-chain-execution-protocol}).
	After that happens, honest participants must append the root of the \emph{latest} graft as soon as it becomes enabled.
	Since this spends $\tx{Init}$, it prevents older grafts to be used.
	Finally, the priority mechanism ensures that this counter-attack always succeeds.
	
	\paragraph{Denial of Service attacks}
	Protocols which make use of timelocks are susceptible to Denial of Service attacks if the adversary is able to delay the actions of participants.
	This threat applies both to the on-chain and off-chain executions.
	Still, several Bitcoin timelock-based protocols are frequently used in practice (\eg, HTLC~\cite{Andrychowicz14sp}, micropayment channels~\cite{Poon15lighting}).
	These protocols are considered to be secure since it is unrealistic for an adversary to be able to completely cut participants off the Bitcoin network for a very long time.
	Using our notation, as long as $\Delta_w$ is chosen to be large enough, we can realistically assume that participants are always able to react in time when needed.
	
	Our off-chain execution protocol also uses a shorter time $\Delta_r$ to detect unresponsive participants.
	Depending on the actual value of $\Delta_r$, a DoS attack that delays a participant for $\Delta_r$ or longer could be realistically feasible.
	Still, we remark that the impact of such attack would not be catastrophic, and can only cause the contract execution to be moved on-chain (and, consequently, fees to be paid).
	The impact is therefore the same as the one caused by a misbehaving participant that stops cooperating.

	\section{On-chain vs off-chain protocols: comparison and assessment}
	We compare our protocol for off-chain execution against the standard on-chain contract execution protocol.
	The main benefits of our new protocol are the following:
	\begin{itemize}
		\item {\bf Only three fees in the best case.} If the contract is brought to completion through off-chain steps, then only three transactions are appended on the blockchain, independently from the size of the original contract. 
		This significantly reduces the cost for deep contracts. 
		\item {\bf No more than two additional fees in any case.}
		Indeed, the worst case is met when some adversary forces the participants to move the execution on-chain after the stipulation phase, \eg, by redeeming $\tx{Head}$ with $\tx{Init}$ immediately.
		In such case, participants only have to pay for the on-chain execution costs, plus two more fees for $\tx{Head}$ and $\tx{Init}$.
		
		Note that if the participants complete at least two off-chain steps, that is already enough to bring the execution cost on par with the original on-chain contract, even if some participant misbehaves on the third step.
		Moreover, any additional off-chain step beyond the second one further reduces the cost of recovering from future adversarial behaviours. 
		\item {\bf No state rollbacks.} After a honest participant completes an off-chain step reaching a contract node, no adversary can put on-chain a transaction corresponding to an earlier node, effectively rolling back one or more execution steps.
		Indeed, after $\tx{Init}$ is appended on-chain, the next transaction to be appended must be the root of some graft. 
		The graft corresponding to the last off-chain step has a shorter timelock \wrt other graft, so the honest participant is able to put its root on-chain before any adversary can do the same with another graft.
		
		\item {\bf No additional waiting in the best case.} 
		With every off-chain step, the timelock on the next graft decreases, reaching 0 when a leaf is reached.
		
	\end{itemize}
	On the other hand, the main drawbacks of off-chain execution are the following:
	\begin{itemize}
		\item {\bf Two additional fees in the worst case.}
		As already mentioned above, in the worst case we do have to pay two more fees.
		
		\item {\bf Redeeming $\tx{Init}$ can require waiting.}
		Except in the best case (where the contract is completed off-chain), the failsafe mechanism requires the grafted subtrees to have timelocks. This means that whenever a participant is forced to move the contract execution on-chain due to some misbehaving participant, they have to wait before they can redeem $\tx{Init}$ with their graft root.
		This is more harmful in the case of an early attack in the first off-chain steps, since the graft timelocks are larger. Also, moving the execution on-chain early will make the contract consume more fees.
		
		\item {\bf Increased number of signatures and messages.}
		At every off-chain step, the participants need to exchange signatures for every transaction in the grafted subtree. 
		Hence, performing many steps requires participants to exchange more signatures compared to the on-chain stipulation and execution phases.
		The worst case scenario happens when the original contract tree has the form of a long chain of  $n$ nodes, each of them having only one child.
		Executing this contract off-chain requires to sign $O(n)$ grafts, each of which contains $O(n)$ transactions on average.
		Overall, this requires each participant to exchange $O(n^2)$ additional signatures.
	\end{itemize}
	
	When considering fees, our protocol provides significant benefits in almost every scenario. The main drawback seems to be due to timelocks, since a malicious participant can delay the completion of a contract by an amount of time that grows linearly with the depth of the contract tree.
	The increased number of signatures does not seem to be too detrimental, especially when considering that for balanced contract trees the overhead is only of $O(n)$ messages.

        \paragraph{Incentivizing honest behaviour}
	We remark that, when the contract leaves redistribute the unspent fees to the participants, all the participants are encouraged to behave honestly.
	This is because forcing the contract execution to be moved on-chain provides no actual benefit for adversaries, but reduces the amount of unspent fees that will be refunded.
	
	\paragraph{Limitations}
	We remark a few limitations of our approach.
	First, participants must always be live and monitor the blockchain for malicious behaviours, reacting to them in a timely fashion. 
	Doing so assumes that adversaries are not able to perform DoS attacks that can stall honest participants for a significant amount of time.
	Indeed, stalling a participant for more than $\Delta_{w}$ can unlock a $\wait{t}$ condition before that participant can react, disrupting the contract execution in both the on-chain and off-chain case.
	For this, we assume that $\Delta_{w}$ is long enough to ensure that honest participants have plenty of time to act, even in presence of DoS attacks or severe congestion of the Bitcoin network.
	Note that this assumption is widespread, since it is applied to all the time-based blockchain protocols, such as hashed time locked contracts \cite{bitcoinsok}, lotteries~\cite{Andrychowicz14sp,BZ17bw}, and micropayment channels~\cite{Poon15lighting,Decker15sss}. Indeed, the security of such protocols is grounded on the assumptions that honest participants can not be prevented to append their transactions in a timely fashion.
	
	By contrast, stalling a participant for the shorter time $\Delta_{r}$ is not as disruptive. In the wort case, that participant can be regarded as unresponsive by the others, prompting them to move the contract execution on-chain. While this is not ideal because we now have to pay more fees, at least the contract semantics is preserved.
	Moreover, if an adversarial participant aims to move the contract on-chain, they can simply stop interacting with others, which is far easier than performing a DoS attack on others.

	Our approach computes the timelocks according to the depth of the original contract tree, which must be finite and statically known. This assumption is shared with the on-chain execution protocol.
	This is inevitable when dealing with the Bitcoin platform, due to the limited expressiveness of its scripting language.
	
	Finally, in our protocol, all the contract fees must be provided in advance, and locked within the $\tx{Head}$ transaction.
	This applies to both the on-chain and off-chain protocols.
	Additionally, we must specify in advance how much each transaction in the tree will pay as fee: this happens at signing time, much sooner than when the fees are actually paid.
	Only after the contract is terminated the unspent fees are refunded to participants. 
	%This problem is also shared by on-chain contract protocols.
	In the off-chain protocol the problem is partially mitigated by the fact that fees could be adjusted when creating the new grafts, if the current market conditions suggest to do so.
	However, the fees in the grafts must still be covered by the $\tx{Head}$ balance, which limits the possible adjustments.
	Further, several hours may pass between the signing of a graft and its actual use, making it harder to accurately predict the appropriate fees.
	Finally, once the execution is moved on-chain fees are again locked.

	\paragraph{Protocol improvement: saving one more fee}
	As discussed above, our protocol is efficient in terms of fees, especially in the best case where all the participants are honest and only \emph{three} fees are needed.
	It is possibly to slightly improve the best case and make it so only \emph{two} fees are needed.
	For that, it suffices to amend the grafting protocol.
	Recall that, when the contract execution is about to end and we create the last graft, involving only a single transaction $\txT$ for a leaf, our protocol signs $\txT$ so that it can redeem $\tx{Init}$.
	After that, we can redeem $\tx{Head}$ with $\tx{Init}$ and then $\tx{Init}$ with $\txT$, closing the contract, having paid a total of three fees. 
	We modify the grafting protocol so that after creating $\tx{T}$ we also create a $\txTi$ with the same outputs, but having $\tx{Head}$ as its input instead of $\tx{Init}$.
	$\txTi$ is then signed accordingly.
	In this way, to close the contract we can directly redeem $\tx{Head}$ with $\txTi$, which involves paying a total of two fees only.

	\paragraph{Protocol improvement: reducing the number of signatures}
    Both the on-chain and off-chain contract stipulation and execution protocols employ a fairly large number of signatures. Indeed, virtually all the transactions require at the very least the implicit signatures, whose number is the same as the number of the participants.
    This impacts on the script size, which should be kept small.
    To improve script size, Bitcoin allows the use of Schnorr signatures~\cite{BIP340}. 
    These allow to combine the signatures of many (cooperating) participants into a single signature. At stipulation time and at grafting time (when the participant cooperate to sign a new graft) participants can communicate to produce a Schnorr signature for the transactions at hand instead of sharing a set of signatures as usual.
    Transactions requiring execution-time authorizations from multiple participants can also benefit from Schnorr signatures, again reducing the number of signatures to a single one.
    Using this technique, we essentially reduce the script of each transaction to be a logical disjunction involving as many cases as the number of children in the contract tree, where each case requires a single Schnorr "implicit" signature, plus another one when authorizations are required by the contract logic.
    Note that we can not combine these two signatures since one is created at stipulation/grafting time, while the other is created at execution time.
    This makes the script size no longer dependent on the number of participants.
    
	\section{Conclusions and related work}
	
	We presented an optimistic protocol for the off-chain execution of Bitcoin smart contracts. Our protocol allows participants to save on transaction fees. In the best case scenario, we only require three fees, while in the worst case we need to pay the cost of the original contract, plus two fees.
	
	Our protocol achieves both security and efficiency by leveraging floating transactions, \ie transactions that are signed off-chain but not yet put on the blockchain.
	More precisely, the security of our protocol is based on the fact that honest participants hold the latest graft floating, and are able at any time to put the graft on chain in order to commit the latest state to the blockchain.
	Instead, the efficiency of our protocol follows from participants not having to put these floating transactions on chain until the very end of the contract (if all participants are honest).
	In general, off-chain protocols (also referred as \emph{layer 2} protocols) are a common solution to improve the scalability of blockchains~\cite{tortola2024layer2,gudgeon2020layer2}.
	
	Our mechanism of floating transactions is reminiscent of the one exploited by the Lightning Network Protocol~\cite{Poon15lighting}.
	The Lightning Network exploits floating transactions to realize \emph{micropayment channels}, allowing participants to transfer bitcoins by simply signing a new floating transactions, without having to append anything on the blockchain and pay the associated fees.
	Such payments are also instantaneous, since there is no need to wait for transactions to be confirmed.
	While such use of floating transactions is a widely adopted and studied approach, so far on stock Bitcoin it has only been used for specific purposes (\eg micropayment channels).
	By contrast, our protocol is general, and can be used to run \emph{any} contract tree off-chain.
	
	We are aware of a few other approaches to improve the execution of Bitcoin contracts.
	For instance, \cite{Decker18eltoo} proposes \textit{eltoo}, a ``layer 2'' protocol to move contracts off-chain.
	This protocol relies on extending Bitcoin with a special signature type (a special \textit{sighash}), which is exploited to allow jumping from an old state to a newer one without having to perform the intermediate steps.
	By contrast, our approach is tailored for Bitcoin as-is, without any extensions.
	
	Another approach~\cite{Das19FASTKITTEN} is to execute contracts off-chain in a controlled way, and then require the result to be certified before it can be put back on the blockchain.
	This approach is feasible but it requires and external trusted execution environment.
	While trusting the hardware is an option, this approach effectively gives the hardware manufacturers the status of a certification authority, and so the power to break contract execution.
	As such, contract are no longer decentralized.
	Our approach instead only relies on basic cryptography, time locks, and the Bitcoin network.

	Smart contracts can be implemented on top of Bitcoin~\cite{Stacks23L2,Nick20liquid} by creating a new ``layer 2'' overlay network with its own consensus mechanism which commits the contract state on Bitcoin.
	This is sometimes referred to as the ``side chain'' approach. 
	Note that, in this setting, the security of the smart contracts no longer relies on the honesty of the majority of the Bitcoin nodes, but also on that of the layer 2 nodes.
	Our approach does not relay on any external complex  infrastructure beyond stock Bitcoin.
	A well-known approach to efficiently run smart contracts in several platforms is to use \emph{rollups}~\cite{TSA22rollupsurvey}
	A first form of rollup is given by \emph{zero-knowledge rollups} \cite{Chaliasos24zkRollups}.
	In this approach, contracts are first executed off-chain, and then their result is put on-chain together with a zero-knowledge proof that the result is indeed the intended one.
	Alternatively, \emph{optimistic rollups} \cite{Kalodner2018ArbitrumSP}  can be used: after the contract is run off-chain, any participant can attempt to put a state on chain and claim it is the correct result.
	After that, other participants can verify the claim and challenge it if they believe the state to be wrong.
	Such challenges must be made within a time window, after which the state becomes final.
	Winning such challenge rejects the proposed new state, and penalizes the proponent.
	Overall, these rollup protocols appear to be very useful but they are designed for powerful platforms like Ethereum where their complex on-chain mechanisms can be implemented.
	On stock Bitcoin, instead, we only have a very limited scripting language, so it is unclear if the approach can be followed.

	\paragraph{Acknowledgments}
	This work was partially supported by project SERICS (PE00000014) under
	the MUR National Recovery and Resilience Plan funded by the European
	Union - NextGenerationEU.
	\bibliography{main}
	
\end{document}

\endinput

